\documentclass[referee]{aa}
\begin{document} 
\thesaurus{11(11.13.1; 11.05.2; 09.16.1)} 
\title{Cosmic Background of Gravitational Waves from Rotating Neutron Stars}
\author{T. Regimbau, J.A. de Freitas Pacheco} 
\offprints{T. Regimbau}
\institute{Observatoire de la C\^ote d'Azur, B.P. 4229, 06304 Nice Cedex 4, France}
\mail{regimbau@obs-nice.fr, pacheco@obs-nice.fr}
\date{Received date; accepted date} 
\maketitle
\markboth{Regimbau.:~Gravitational Background}{}

\begin{abstract}
{The extragalactic background of gravitational waves produced
by tri-axial rotating neutron stars was calculated, under the assumption
that the properties of the underlying pulsar population are the same of
those of the galactic population, recently derived by Regimbau \& de Freitas
Pacheco (2000). For an equatorial ellipticity of $\varepsilon$ = 10$^{-6}$,
the equivalent density parameter due to gravitational
waves has a maximum amplitude in the range 2$\times$10$^{-11}$-3$\times$10$^{-9}$, around 
0.9-1.5 kHz. The main reasons affecting the theoretical predictions are discussed. This
background is comparable to that produced by the ''ring-down'' emission from
distorted black holes. The detection possibility of this background by a future 
generation of gravitational antennas is also examined.}

\end{abstract} 


\section{Introduction}

In the past years, a large number of papers devoted to stochastic
 backgrounds of gravitational waves appeared in the literature (see
Maggiore 2000 for a recent review). Besides processes that took place
very shortly after the big-bang, the emission from a large number
of unresolved sources can produce a stochastic background. Supernovas
(Blair et al. 1997) and distorted black holes (Ferrari et al. 1999a; 
de Ara\'ujo et al. 2000) are examples of sources able to generate a shot 
noise, while a truly continuous background could
be produced, for instance, by the '' r-mode'' emission from young and
hot neutron stars (Owen et al. 1998; Ferrari et al. 1999b).
Detection of such backgrounds may probe the cosmic star
formation rate up to redshifts of z $\sim$ 4-5,
the mass range of the progenitors of neutron stars and black holes as well as
the initial angular momentum of these objects.

The contribution of the entire population of rotating neutron stars to the
 continuous
galactic background of gravitational waves was considered by different authors
(Schutz 1991; Giazotto et al. 1997; de Freitas Pacheco \& Horvath 1997)
 and, more
recently, this subject was revisited by Regimbau \& de Freitas Pacheco
 (2000, hereafter
RP00). In the latter, the ``true'' population of rotating neutron stars
 was synthesized by
Monte Carlo techniques and its contribution to the galactic background
 of gravitational waves
was estimated. If the planned sensitivity of the first generation of laser
 beam interferometers
is taken into account (in particular that of the French-Italian
 project VIRGO), then the
simulations by RP00 suggest that only few objects will contribute to the
 signal, if the
mean equatorial ellipticity of  neutron stars is of the order
 of $\varepsilon$ = 10$^{-6}$.
Upper limits on $\varepsilon$ have been obtained by  assuming that  the
 observed spin-down
of pulsars is essentially due to the emission of gravitational waves. In
 this case, one obtains
$\varepsilon \leq$ 10$^{-3}$ for ``normal'' pulsars whereas recycled or
 rejuvenated
pulsars seem to have equatorial deformations less than 10$^{-8}$.
Although the galactic population will not produce a truly background, it
remains to be investigated the integrated contribution of these objects
in a fairly large volume of the universe. Estimates of this emission are important
because it may rival with a possible background of cosmological origin. Since
there is an upper limit to the wave frequency of the pulsar gravitational
radiation, a putative cosmic background will dominate at low frequencies
and, the knowledge of the spectral energy distribution of the background
produced by discrete sources may help in the choice of the best frequency
domain to search for a relic emission. 

In the present paper, the integrated gravitational emission of rotating
neutron stars to the background is calculated under the assumption that 
the distributions of the rotation period and magnetic field
derived by RP00 can conveniently be scaled to other galaxies. 
The plan of this paper in the following: in
section 2 the model computations are described; in section 3 
the results are discussed and finally, in section 4 the conclusions are given.

\section{The Model}

The main working hypothesis of  our computations concerns the true 
rotation period and magnetic field distributions of pulsars. For galactic 
objects, these distributions were derived by RP00 using
Monte Carlo simulations to reproduce the different observed distributions of
physical parameters, like the period and its first derivative as well as distances,
when selection effects are taken into account. These simulations 
permitted to establish the parameters of the initial distribution of period 
and magnetic field. ``Faute de mieux'', we assumed here that pulsars are born 
everywhere with rotation periods and fields obeying the same distribution laws.

For a single pulsar, the frequency distribution of the total emitted 
gravitational energy in the source's frame is
\begin{equation}
{{dE}\over{d\nu}} = {{dE}\over{dt}}\left\vert{{dt}\over{d\nu}}\right\vert
\end{equation}
It is worth mentioning that in spite of (dE/dt) in the above equation be the pulsar
gravitational wave emission rate, the time variation of the frequency is
fixed by the magnetic dipole emission, responsible for the deceleration of
the star. This means that angular momentum losses by gravitational waves will never
overcome those produced by magnetic torques, which is equivalent to say that
the average equatorial deformation is always less than 10$^{-3}$. 
Under these conditions, the energy frequency distribution is
(the fact that the gravitational wave frequency is twice the rotational
frequency was already taken into account)
\begin{equation}
{{dE}\over{d\nu}} = {{256G\pi^6}\over{5c^5}}\varepsilon^2I^2({{\tau_m}\over{P_o^2}})
\nu^3 = K\nu^3
\end{equation}
where I is the moment of inertia of the star, $\tau_m$ is the magnetic braking
timescale (see RP00), P$_o$ is the initial period of the pulsar and the other
symbols have their usual meaning. 

In order to estimate the average ratio $(\tau_m/P_o^2)$, we adopted the following
procedure. We have performed Monte Carlo simulations in which the distribution
probabilities of the variables $\tau_m$ and P$_o$ are the same as RP00.
The resulting distribution of the quantity log$(\tau_m/P_o^2)$ is given in fig.1, and it
can be fitted by a Gaussian with a mean equal to $<$log$(\tau_m/P_o^2)>$ = 12.544.
Adopting this value as representative of the whole population, 
the constant K in eq. (2) is K = 9.27$\times$10$^{35}(\varepsilon_{-6})^2$ 
\, erg.Hz$^{-4}$, where we have introduced the notation $\varepsilon_{-6} = 
(\varepsilon/10^{-6})$.

The gravitational wave flux at frequency $\nu_o$ (observer's frame) due to 
sources localized in the redshift shell z,z+dz is
\begin{equation}
dF_{\nu_o} = {{1}\over{4\pi d_L^2}}{{dE}\over{d\nu}}\left\vert {{d\nu}\over
{d\nu_o}}\right\vert dR(z) 
\end{equation}
where $d_L$=(1+z)r is the distance-luminosity, r is the proper distance and the
observer's frequency $\nu_o$ is related to the frequency $\nu$ at the source by
$\nu = (1+z)\nu_o$. The event rate inside the shell z,z+dz is
\begin{equation}
dR(z) = \lambda_pR_c(z){{dV}\over{dz}}dz
\end{equation}
In the above equation, R$_c(z)$ is the ''cosmic'' star formation rate, 
$\lambda_p = \int^{40}_{10}\xi(m)dm$ is the mass fraction of formed
stars in the range 10-40 M$_{\odot}$, supposed to be the mass range
of the pulsar progenitors, with $\xi(m)$ being the initial mass function.
For a Salpeter's law ($\xi(m) \propto m^{-2.35}$), $\lambda_p$ = 4.84$\times
10^{-3}$ M$_{\odot}^{-1}$.
The element of the comoving volume is
\begin{equation}
dV = 4\pi r^2{{c}\over{H_o}}{{dz}\over{E(\Omega_i,z)}}
\end{equation}
where H$_o$ is the Hubble parameter and the function E$(\Omega_i,z)$ is
defined by the equation
\begin{equation}
E(\Omega_i,z) = \left\lbrack(1+z)^2(1+z\Omega_m)-z(2+z]\Omega_v\right\rbrack^{1/2}
\end{equation}
where $\Omega_m$ and $\Omega_v$ are respectively the density parameters
due to matter (baryonic and non-baryonic) and the vacuum, corresponding
to a non-zero cosmological constant. The equivalent
density parameter due to the spatial curvature satisfies $\Omega_k = 1-\Omega_m -\Omega_v$.  

Combining these equations, the expected gravitational wave flux at the
frequency $\nu_o$ is
\begin{equation}
F_{\nu_o}= K\nu_o^3\lambda_p({{c}\over{H_o}})(\varepsilon_{-6})^2\int_0^{z_{max}}
{{(1+z)^2R_c(z)}\over{E(\Omega_i,z)}}dz
\end{equation}
In the literature is often used an equivalent density parameter due to gravitational waves 
to measure the strength of the background at a given frequency, defined by the equation
\begin{equation}
\Omega_{GW}= ({{8\pi G}\over{3H_o^2}}){{\nu_oF_{\nu_o}}\over{c^3}}
\end{equation}

\subsection{Numerical calculations}

In order to evaluate numerically eqs. (7)-(8), it is necessary to specify the 
 cosmic star formation rate R$_c(z)$ and the parameters of the world
model, namely, the values of H$_o$, $\Omega_m$ and $\Omega_v$.

Madau \& Pozzetti (1999) have reviewed the constraints imposed by the
observed extragalactic background light on the cosmic star formation rate (CSFR). They 
concluded that after an extinction correction of A$_{1500}$ = 1.2 mag 
(A$_{2800}$ = 0.55 mag), a star formation rate given by the relation
\begin{equation}
{R_c(z)} = {{0.23e^{3.4z}}\over{(44.7 + e^{3.8z})}} \,\, M_{\odot}.yr^{-1}.Mpc^{-3}
\end{equation}
fits well all measurements of the UV-continuum and H$\alpha$ luminosity densities
from the present epoch up to z = 4. However, according to Hopkins et al. (2001),
even when reddening corrections are taken into account, significant discrepancies
still remain between the CSFR derived from UV-H$\alpha$ measurements and those derived
from far-infrared and radio luminosities, which are not affected by dust extinction.
Hopkins et al. (2001) assumed a reddening correction {\it dependent on} the star formation
rate and obtained a good agreement between the CSFRs derived from different set of
measurements. We have fitted their results by a function similar to eq.(9), namely,
\begin{equation}
{R_c(z)} = {{1.207e^{3.836z}}\over{(39.970 + e^{4.163z})}} \,\, M_{\odot}.yr^{-1}.Mpc^{-3}
\end{equation}
Taking into account the uncertainties still present in the derivation of the CSFR,
we have performed  calculations  using both rates.

Recent BOOMERANG and MAXIMA results (de Bernardis et al. 2000; Hanany et al. 2000) on
the power spectra of the cosmic microwave background and observations of distant type 
Ia supernovas (Perlmutter et al. 1999; Schmidt et al. 1998), which suggest that the 
expansion of the Universe is accelerating, support a spatially flat geometry and
a non-zero cosmological constant. Both set of data are consistent with $\Omega_m$ = 0.30
(including baryonic and non-baryonic matter) and $\Omega_v$ = 0.70, which will be adopted
in our computations. However, no significant differences in our results were
observed if a model defined by $\Omega_m$ = 1 and $\Omega_v$ = 0 is adopted. The
Hubble parameter H$_o$ was taken to be equal to 68 km/s/Mpc (Krauss 2001). 

RP00 assume in their simulations that the maximum rotation frequency of a newly
born pulsar is 2000 Hz, which corresponds to a gravitational wave frequency of
4000 Hz. If the upper limit of the integral in eq. (7) is z$_{max}$ = 5, then
the maximum frequency seen by the observer is $\approx$ 660 Hz. For higher
frequencies, only near objects will contribute to the integral and the upper
limit should be replaced by $z_{max}=(4000/\nu_o)-1$, with $\nu_o$ in Hz. This
parameter affects the resulting spectrum as we shall see below. Thus, calculations
with a different cutoff were also performed.

\section{Results}

Figure 2 shows the density parameter $\Omega_{GW}$ as a function of the frequency.
Labels M1 and H1 correspond to star formation rates given respectively
by eqs. (9) and (10) and a maximum gravitational wave frequency equal to 4000 Hz.
The labels M2 and H2 have the same meaning but here
the maximum gravitational wave frequency cutoff is at 2000 Hz,
corresponding to a minimum rotation period of 1ms. All these curves were
calculated for an equatorial deformation $\varepsilon$=10$^{-6}$, and we
recall that the
results scale as $\varepsilon^2$. The numerical calculations indicate a broad 
maximum around 1.5 kHz, if the maximum possible pulsar rotation frequency 
is 2 kHz. Decreasing this limit by a half, the spectrum narrows and the maximum 
shifts toward lower frequencies ($\sim$ 0.9 kHz). The amplitude is also affected, being
reduced by almost one order of magnitude. This happens because according
to eq.(8), the amplitude grows as $\nu_o^4$ but a lower frequency cutoff
implies that only nearby objects will contribute to the integrated signal and
thus reducing the amplitude at maximum.

If the equatorial ellipticity may reach values of the order of 10$^{-6}$,
then the energy density of the background generated by pulsars may be comparable
and even higher than that expected from newly born black holes (Ferrari et al.
1999a; de Ara\'ujo et al. 2000), originated from the collapse of massive stars
(M $\geq$ 40 M$_{\odot}$). For a comparison, the spectrum corresponding to
the ring-down emission from distorted black holes calculated by Ferrari et
al. (1999a) is also plotted in fig.(2), appropriately scaled to the Hubble
parameter here adopted. In the case of distorted black holes, the
uncertainties on the estimates of the background energy density rest on 
the conversion efficiency of the mass energy into gravitational waves as well as on
the minimum mass of the progenitor. 

Hot and fast rotating newly formed
neutron stars may be unstable against the r-mode instability. Ferrari
et al. (1999b) estimated that if all newly born neutron stars cross
the ''instability window'' (see, for instance, Andersson et al. 2000), then
the resulting density parameter has a maximum amplitude of 
$\Omega_{GW} \approx 2\times 10^{-8}$ in the frequency range 0.5-1.7 kHz. This
signal by far would be the dominant component of the background at these frequencies.
However, according to the simulations by RP00, only few pulsars are born
within the instability window, reducing the amplitude of
the background due to such a mechanism by orders of magnitude.

Unless the equatorial ellipticity of pulsars be substantially higher than
the present expectations, the background generated by rotating neutron
stars will hardly be detected by the present generation of laser
beam interferometers and/or resonant detectors, but this could be a possibility
for future projects presently under consideration, as the Large
Scale Cryogenic Gravitational Wave Telescope (LCGT), sponsored by the University
of Tokyo and the European antenna EURO (W. Winkler, private communication). The former,
with a baseline of 3 km, is expected to have 
a 100W laser and cooled sapphire mirrors among other technological improvements.
Therefore, one may expect that advanced laser beam interferometers may attain   
in a near future a sensitivity around 1 kHz, corresponding to
a strain noise $\sqrt{S_n(\nu)}$ of about 10$^{-25}$ Hz$^{-1/2}$. On the other hand,
the best strategy to detect the signal, when the detector output is
dominated by the noise, which is the present case, is to correlate data from two 
different gravitational antennas and to assume that they have independent noise. 
One interesting possibility would be to correlate the output of such an advanced
detector with a resonant mass detector located at the same site, having a spherical 
or truncated icosahedron geometry. The advantages of this geometry
with respect to a longitudinal bar is that a free elastic sphere has five
degenerate quadrupole modes, each of which is sensitive to a different 
polarization and wave direction. Moreover, for a given material and resonant
frequency, a spherical detector has a cross section larger than a cylindrical one. The
sensitivity of resonant spheres is limited by Brownian motion noise associated 
with dissipation in the antenna and transducer, as well as by the electronic
noise from amplifiers. In this case, the strain noise at resonance is approximately 
(Coccia \& Fafone 1997) 
\begin{equation}
\sqrt{S_n(\nu)} = ({{4kT}\over{F_nQ_nM_sv_s^2\nu_n}})^{1/2}
\end{equation}
where k is the Boltzmann constant, T is the sphere temperature, F$_n$ is
a dimensionless coefficient depending on each quadrupole mode (F$_1$ =
2.98, F$_2$ = 1.14, F$_3$ = 0.107), Q$_n$ is the quality factor of the mode,
M$_s$ is the mass of the sphere, v$_s$ is the 
velocity of the sound and $\nu_n$ is the mode frequency. For practical purposes,
let us consider a sphere constituted of the aluminium alloy Al-5056. This
material has a sound velocity of 5440 m.s$^{-1}$ and Coccia et al. (1996)
have reported Q values as high as 10$^8$ for temperatures below 100 mK. A sphere
with a diameter of 3.5 m (mass of 60.3 tons) has the two main frequencies
of the quadrupole modes at 0.8 kHz and 1.5 kHz,
covering quite well the predicted interval where the maximum amplitude of the
pulsar background should occur. Assuming a typical temperature of 20 mK, the expected 
strain noise derived from eq.(11) is
$\sqrt{S_n(\nu_1)} \approx$ 1.5$\times$10$^{-24}$ Hz$^{-1/2}$.

If $\Delta\nu \approx$ 20 Hz is the bandwidth of the resonant mass detector and
$\tau$ is the integration time, then the expected optimized signal-to-noise S/N of 
the correlated outputs is (Allen 1997)
\begin{equation}
{{S}\over{N}} = {{3H_o^2}\over{\sqrt{50}\pi^2}}\sqrt{\Delta\nu \tau}{{\Omega_{GW}}
\over{\nu^3\sqrt{S_1(\nu)S_2(\nu)}}} 
\end{equation}
For one year integration, one obtains from the equation above S/N $\approx$ 0.2,
indicating that new technology detectors may reach in the future the required
sensitivity to detect such a signal.

\section{Conclusions}

The contribution of rotating neutron stars to the extragalactic background of
gravitational waves was calculated, under the assumption that the parameters
characterizing the galactic population of pulsars derived by RP00
are the same everywhere.
The amplitude of the equivalent density parameter attains a maximum in
the frequency interval 0.9-1.5 kHz and is in the range 10$^{-11}$ up to
3$\times$10$^{-9}$. The amplitude scales as $\varepsilon^2$ and, for a given equatorial 
ellipticity, the main uncertainties in the amplitude are essentially due to the
cosmic star formation rate and to the rotation frequency limit at the
pulsar birth, which depends on the equation of state of the nuclear matter. For
''realistic'' equations of state, these limits are in the rotation period range 0.5 - 1.0
ms, values adopted in our calculations.

The present estimates indicate that this background, having a duty cycle
(measured by the product between the typical duration of the signal and the
mean birth frequency of pulsars) greater than one, may have an energy density 
comparable to that produced by ''ring-down'' black holes. This emission is unlikely to be
detected by the present generation of detectors. Correlated advanced detectors 
may reach a limit of about $\Omega_{GW} \approx$ 10$^{-10}$ 
for a {\it flat} spectrum (Maggiore 2000), which is not the present case. However,
new technology detectors, which are presently under consideration, may
attain the required sensitivity. In particular, taking into account the
low cost of a resonant mass detector when compared with that of a laser interferometer,
the installation in the same site of a ''sphere'' operating near the
maximum predict frequency ($\sim$ 0.9-1.5 kHz), could be the adequate strategy
to detect such a signal in the future.

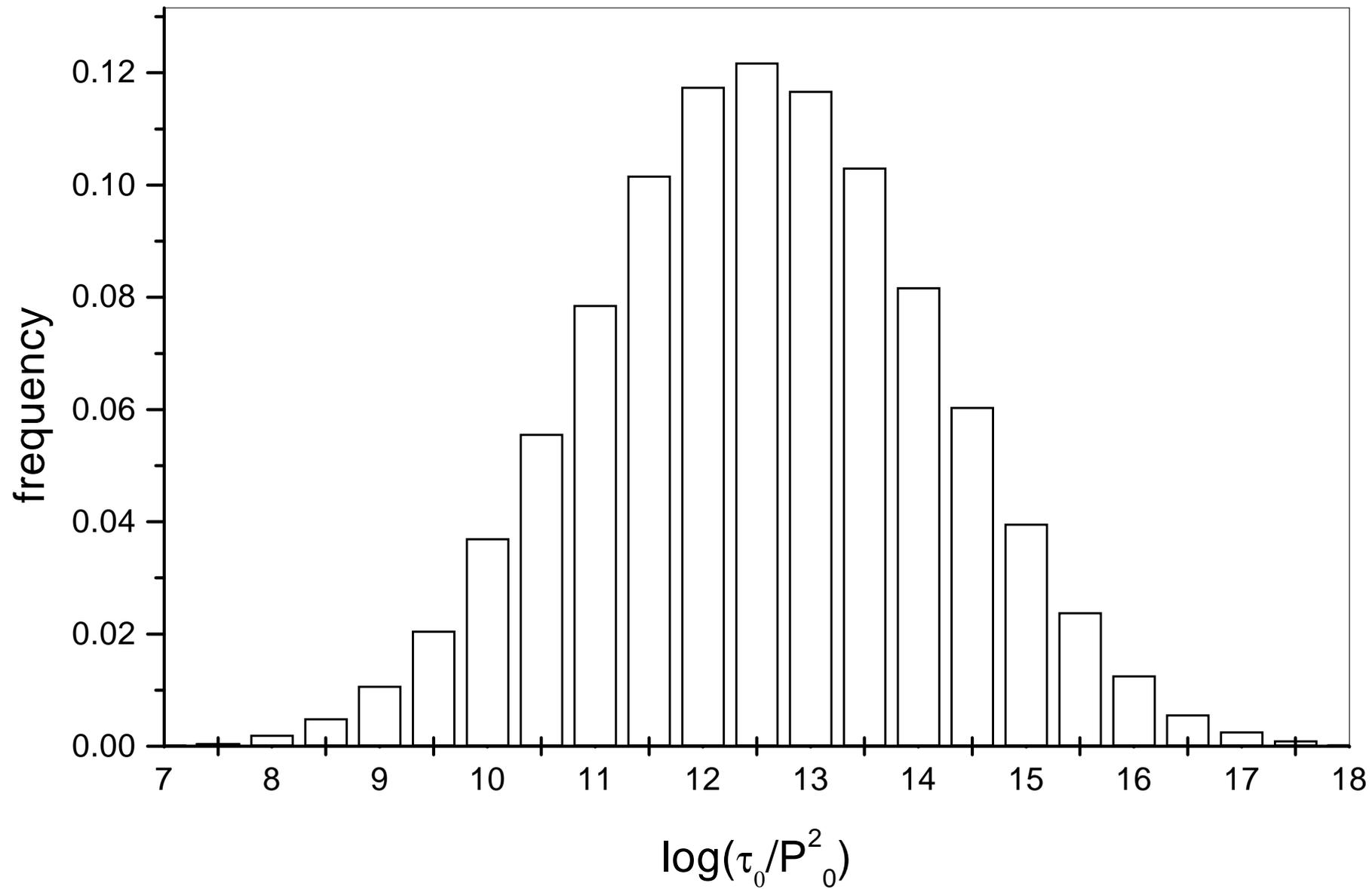
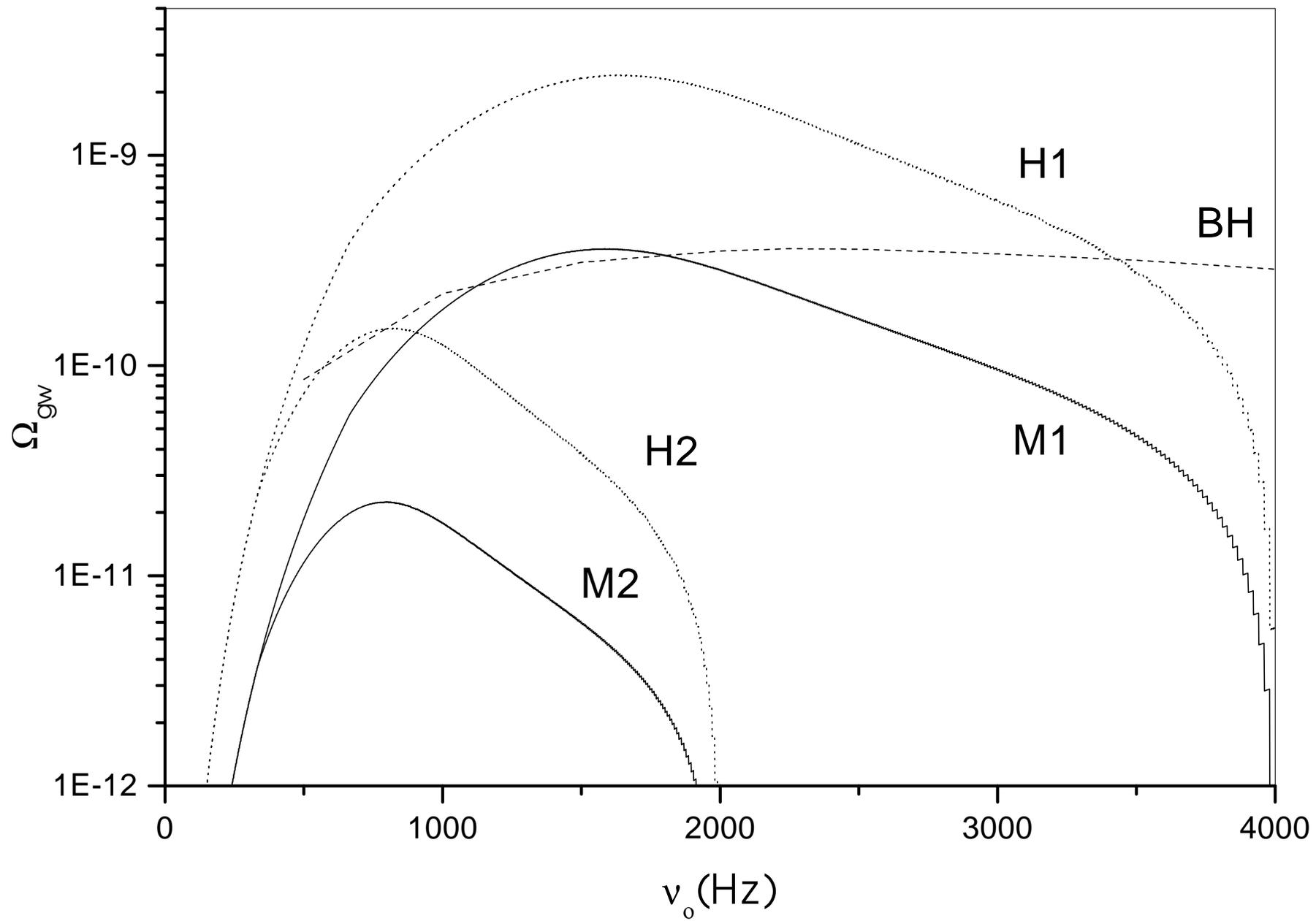
\end{document}